\documentclass[prl,aps,twocolumn,showpacs,preprintnumbers,amsmath,amssymb]{revtex4}
\usepackage{epsf}

\begin{document}
\draft
\newcommand{\ve}[1]{\boldsymbol{#1}}

\title{Interface hole-doping in cuprate-titanate superlattices}
\author{N.~Pavlenko$^1$\cite{address2}, I.~Elfimov$^2$, T.~Kopp$^1$, and G.A.~Sawatzky$^2$}
\address{$^1$EKM, Universit\"at Augsburg, 86135 Augsburg, Germany\\
$^2$Department of Physics and Astronomy, University of British Columbia, Vancouver, Canada V6T1Z1}

\begin{abstract}
The electronic structure of interfaces between YBa$_2$Cu$_3$O$_6$ and SrTiO$_3$
is studied using local spin density approximation (LSDA) with intra-atomic
Coulomb repulsion (LSDA+U). We find a metallic state in cuprate/titanate
heterostructures with the hole carriers concentrated substantially in the
CuO$_2$-layers and in the first interface TiO$_2$ and SrO planes. This
effective interface doping appears due to the polarity of interfaces, caused by
the first incomplete copper oxide unit cell. Interface-induced
high pre-doping of CuO$_2$-layers is a key mechanism controlling the
superconducting properties in engineered field-effect devices realized on
the basis of cuprate/titanate superlattices.
\end{abstract}

\pacs{74.81.-g,74.78.-w,73.20.-r,73.20.Mf}

\maketitle

In complex thin-film oxide heterostructures of structurally compatible but
physically dissimilar compounds, interface phenomena can substantially affect
the charge properties. A prominent example \cite{ohtomo} is the titanate
superlattice composed of insulating layers of SrTiO$_3$ and LaTiO$_3$, where
the mixed valence ($+3$/$+4$) of Ti leads to an interface-driven electronic
redistribution and to metallic conductivity. Moreover, when one of the
superlattice compounds is a copper oxide film, where the high-$T_c$ properties
can be tuned by doping, the behavior is even more intriguing. Such
heterostructures consisting of YBa$_2$Cu$_3$O$_{7-\delta}$-films grown on
SrTiO$_3$-layers are of essential importance due to their applications in
superconducting field effect devices \cite{mannhart}.
It is well established that external electrostatic fields can significantly
affect the superconducting transition temperature ($T_c$) in these layered
materials which is often understood in terms of electrostatic doping
[the more charge is field-injected into the film the larger $T_c$]\cite{mannhart,logvenov}.
Despite capturing the key mechanism of charge
modulation in the field-effect, this concept does not include a detailed
consideration of the microstructure of the YBa$_2$Cu$_3$O$_{7-\delta}$-film
near the interface assuming that the latter remains unaffected by the adjacent
SrTiO$_3$ interface layer.

Several experimental facts, however, indicate an interface-related change of
the electronic states in the cuprate/perovskite oxide heterostructures. First,
recent experimental studies performed on the underdoped cuprate films resulted
in a $T_c$-shift of about 5--15~K, whereas in the overdoped films no shifts
were observed, a fact, which cannot be  explained satisfactorily by
field-doping \cite{mannhart}. Furthermore, studies of hole mobility in
the CuO$_2$ planes of SrTiO$_3$-cuprate superlattices suggest a substantial
localization of injected holes even above the hole-density level necessary for
a bulk superconductor-insulator transition \cite{eckstein,pavlenko_kopp}.
However, little is known about the electronic properties of the interfaces
between the copper and titanium oxides. This is even more surprising
considering the fact that despite the different physical properties, the
structural compatibility of the cubic SrTiO$_3$ and YBa$_2$Cu$_3$O$_{7-\delta}$
makes them good candidates to assemble heterostructures and study interfacial
phenomena. The electronic band insulating state of bulk SrTiO$_3$ with a wide
gap of about $3$~eV between the valence O $2p$ band and empty Ti $3d$ bands is
reasonably well described within
an LSDA approach. In contrast, the standard band
theory calculations fail to describe the antiferromagnetic Mott insulating
state of strongly underdoped or undoped cuprates like YBa$_2$Cu$_3$O$_6$.
Instead, the band properties of YBa$_2$Cu$_3$O$_6$ with a gap of about $1.5$~eV
determined by Cu $3d$ and oxygen $2p$ electrons can be satisfactorily treated by introducing the
intra-atomic orbital dependent Coulomb repulsion for the electrons in Cu $3d$ orbitals.

To provide deeper insight into the interface physics of such heterostructures,
we present and interpret results of electronic structure calculations for the
superlattice based on insulating YBa$_2$Cu$_3$O$_6$-films and SrTiO$_3$. As
SrTiO$_3$ consists of an alternating sequence of electrostatically neutral
(001) layers, one can expect that in such heterostructures, the chemical
bonding at the (001) interface with YBa$_2$Cu$_3$O$_6$ will be determined by
the first termination layer which can be either SrO or TiO$_2$. It is worth
pointing out that if the first unit cell of a YBa$_2$Cu$_3$O$_6$-film were
completely grown on the surface of SrTiO$_3$, the interface would be
electrostatically neutral. In this case,
% our LSDA+$U$ calculations
%performed for antiferromagnetic arrangement in CuO$_2$ planes
%produce an insulating state with an
%energy gap of $1.2$~eV between the Cu~$d_{x^2-y^2}$ and O~$p_x$ and $p_y$
%orbitals, similar to the bulk YBa$_2$Cu$_3$O$_6$. Then
the direct influence of the interface would be essentially reduced to the small
change in the band structure of YBa$_2$Cu$_3$O$_6$, originating from the
mismatch of the lattice constants ($a=3.9~$\AA\ in cubic SrTiO$_3$ versus
$a=3.86$~\AA\ in YBa$_2$Cu$_3$O$_6$). However, the recent X-ray studies of
interface arrangement \cite{bals,abbamonte} show clear indications of the
incompletely growing unit cells of cuprate films at the SrTiO$_3$ substrate,
which may drastically change the interface electronic properties.

\begin{figure}[t]
\epsfxsize=7.8cm {\epsffile{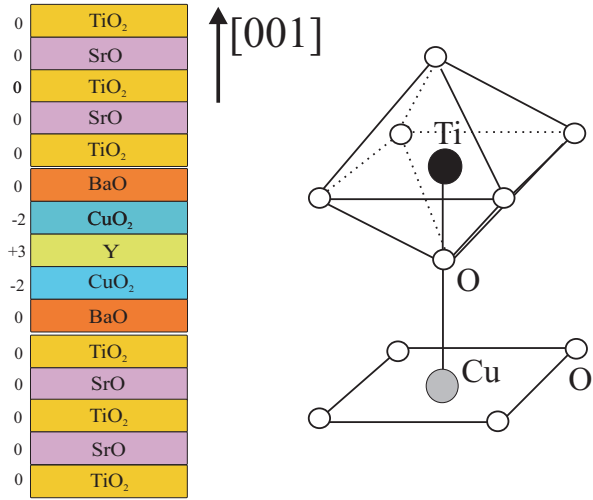}} \caption{Scheme of a
YBaCuO/SrTiO$_3$-sandwich where the polar interfaces appear due to the
incomplete Ba$_2$Cu$_2$O$_6$-unit cell with an interface structural
configuration shown in detail on the right panel.} \label{fig1}
\end{figure}

In this case the interface between polar layers, formed with the anisotropic
YBa$_2$Cu$_3$O$_6$-crystal structure, and non-polar SrTiO$_3$
(001) planes would
result in the so called ``polar catastrophe'' which appears on account of the
divergent electrical potential \cite{wolf}.
The electronic compensation of
the divergence can be achieved by a redistribution of the extra charge carriers
near the interface which leads to a dramatical change of the electronic states
in such a heterostructure. The possibility of such an electronic reconstruction has
been demonstrated by Hesper {\it et al.} for a polar (111) surface of
K$_3$C$_{60}$ \cite{hesper}.

To illustrate the resulting electronic properties, we consider first a
superlattice formed on the basis of a sandwich-type supercell. The
YBaCuO/SrTiO-supercell shown in Fig.~\ref{fig1} consists of an incomplete
copper oxide unit cell YBa$_2$Cu$_2$O$_6$ shared between two layers, each
containing 2 unit cells of SrTiO$_3$. Effectively, the interface bonding here
appears by a `substitution' of the CuO chains, terminating a full
YBa$_2$Cu$_3$O$_{7-\delta}$ cell, with the TiO$_2$-planes which is illustrated
in the right panel of Fig.~\ref{fig1}. Our choice of interface bonding is
strongly motivated by recent TEM-studies of the YBa$_2$Cu$_3$O$_{7-\delta}$
films and similar compounds grown on SrTiO$_3$ with  pulsed laser deposition
technique \cite{bals}. In the case, when the substrate of SrTiO$_3$ is
terminated by a TiO$_2$-layer, the determined interface bonding arrangement is
typically a stack of
$\ldots$/SrO/TiO$_2$/BaO/CuO$_2$/Y/CuO$_2$/BaO/CuO/$\ldots$ layers. Such
structural stacks suggest an interface chemical bonding Ti-O-Cu with the oxygen
of the BaO-layers shared between the CuO$_2$ and TiO$_2$-planes.
From the electrostatical point of view, the initial `bulk-type'
electronic charging of the constituent layers indicated in the left panel of
Fig.~\ref{fig1} would result in 1 extra hole which is needed in order to
compensate the polarity.
From the point of view of symmetry, this compensation leads to a doping of each
block $\ldots$/SrO/TiO$_2$/BaO/CuO$_2$ by 0.5 hole.

To understand the redistribution of the extra charge density, which would
appear near the interface, we calculated the densities of states of a
YBa$_2$Cu$_2$O$_6$/SrTiO$_3$ sandwich using the linearized augmented plane wave
method (LAPW) implemented in the WIEN2k package \cite{wien2k}. Technical
details include the SIC-variant of the LSDA+$U$ method \cite{anisimov} on a
$9\times 9 \times 1$ $\ve{k}$-point grid with $U=8$~eV and $J=0.8$~eV on the Cu
$3d$ orbitals. The lattice constants $a=b=3.8984$~\AA\ are fixed to the
structural values of SrTiO$_3$, whereas the optimized interface distance
$\Delta=1.85$~\AA\ between the apical oxygens of BaO layers and TiO$_2$ planes
has been found by minimization of the total energy, which corresponds to
$c=27.53$~\AA.

\begin{figure}[t]
\epsfxsize=8.5cm {\epsffile{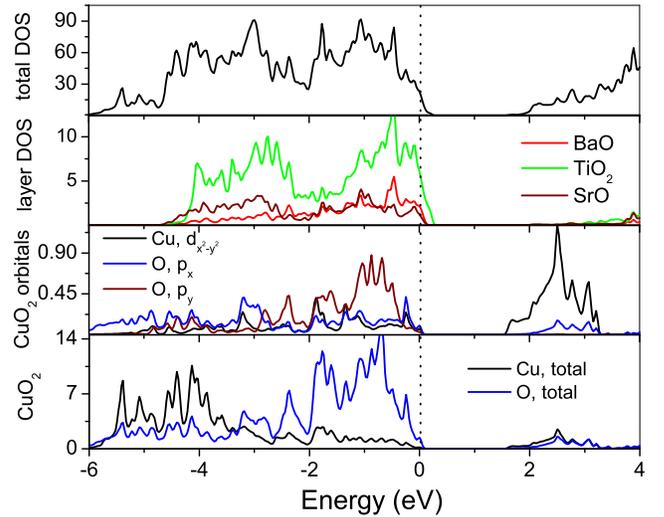}} \caption{Density of states of
the SrTiO$_3$/YBa$_2$Cu$_2$O$_6$/SrTiO$_3$-sandwich calculated within LSDA+$U$
approach with $U=8$~eV and $J=0.8$~eV for the electrons in Cu $3d$ orbitals.
The zero of energy is at Fermi level.} \label{fig2}
\end{figure}
\begin{figure}[t]

\epsfxsize=8.5cm {\epsffile{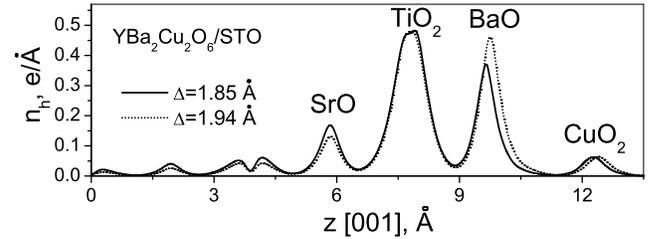}} \caption{Hole density
distribution near the YBa$_2$Cu$_2$O$_6$/SrTiO$_3$-interface for optimized
($\Delta=1.85$~\AA) and unrelaxed ($\Delta=1.94$~\AA) cases. The position $z=0$
is at bottom TiO$_2$-plane of SrTiO$_3$/YBa$_2$Cu$_2$O$_6$/SrTiO$_3$-sandwich.}
\label{fig3}
\end{figure}

Fig.~\ref{fig2} shows the calculated density of states where the position of
the Fermi level $E_F$ is indicated by dots. One can immediately identify the
metallic state with hole carriers in the superlattice from the total density of
states as originating from the oxygen $p$ states. We note that, similar to bulk
YBa$_2$Cu$_2$O$_6$, the Cu~$d_{x^2-y^2}$ states are empty and separated by a gap of
$1.34$~eV from oxygen 2$p$ whereas Cu $d_{3z^2-r^2}$ and $t_{2g}$ bands remain
below the Fermi level. As one can see from Fig.~\ref{fig2}, a substantial amount
of charge compensating hole density is distributed over CuO$_2$ planes.
However, we find that also BaO layers as well as the first interface TiO$_2$
and SrO planes are doped. The upper boundaries of the O $p$ bands of the more
distant SrO and TiO$_2$ planes (with respect to the interface) remain almost on
the same level with $E_F$ which implies that the charge is confined essentially
in the interface block of SrO/TiO$_2$/BaO/CuO$_2$ layers. Fig.~\ref{fig3} shows
the distribution of hole density spatially resolved along $z$ ([001])-direction
within this interface block and calculated for optimized ($\Delta=1.85$~\AA)
and unrelaxed ($\Delta=1.94$~\AA) sandwiches. To obtain this quantity, we have
generated the charge density in the energy interval between the Fermi level and
the top of the valence band. Specifically, we obtain that, although
approximately $5\%$ of the hole density is located in the CuO$_2$ planes, the
major part is concentrated within the BaO ($\approx 25\%$) and the first
TiO$_2$ ($48\%$) and SrO ($12\%$) layers. This suggests a finite metallic
conductivity in the titanate, BaO and copper oxide planes. Furthermore, while
the relaxation of the structure leads to the reduction of the hole density in
the BaO plane and to its redistribution within the interface SrO and more
distant planes, the hole density in the CuO$_2$ planes remains almost
unaffected.

\begin{figure}[t]
\epsfxsize=7.8cm {\epsffile{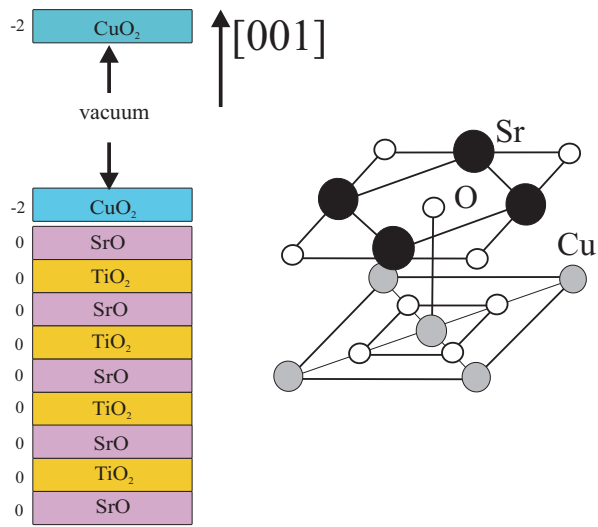}} \caption{Scheme of a polar
CuO$_2$/SrTiO$_3$-superlattice where a STO-layer is terminated by a SrO-plane.
The right panel shows a structural configuration which appears at the
interface.} \label{fig4}
\end{figure}

The resultant hole charge in the copper oxide planes of cuprate/titanate
heterostructures clearly demonstrates that, apart from chemical doping, the
interface polarity is another important mechanism which modulates the doping
level in the cuprate films. In superconducting field effect devices, operated
by electrostatic charging, an initial pre-doped hole density, caused by
the interface, may have striking consequences on
their performance. The most important feature is the $T_c$ shift which may be
directly affected when a typically achievable hole density $\sim 0.05$ is
injected into the already pre-doped ($x=0.025$) copper oxide film. Moreover, it
appears that much higher hole doping levels including strong overdoping at the
interfaces can be obtained in other interface configurations when the first
unit cell of YBa$_2$Cu$_3$O$_{7-\delta}$ at the interface remains incomplete.

To demonstrate such a structurally induced overdoping, we consider a case in
which a copper oxide plane is directily deposited on a SrTiO$_3$ substrate
terminated by SrO (Fig.~\ref{fig4}). The direct deposition of the single
Cu$^{2+}$O$_2^{4-}$ plane on the non-polar titanate layer would require 2 extra
holes to maintain the overall charge neutrality. To achieve such an extremely
high doping level, interface electronic reconstruction is inevitably required.
The importance of electronic reconstruction is strongly supported by
significant smoothing of interference fringes observed with anomalous X-ray
scattering in doped La$_2$CuO$_{4+\delta}$-films---an effect explained by the
mobile carrier depletion in cuprate films near the SrTiO$_3$ substrate
\cite{abbamonte}. Apart from the electronic mechanism, other forms of interface
reconstruction could modify the chemical composition. For example, the ionic
compensation due to the cation intermixing and oxygen vacancies (oxygens
missing in CuO$_2$ during the growth) will be a competing mechanism to
compensate the polarity \cite{wolf}. However, it is still instructive
to enforce atomically flat and stochiometric surfaces in order to investigate
comprehensively the electronic mechanism \cite{multi}.

\begin{figure}[t]
\epsfxsize=8.5cm {\epsffile{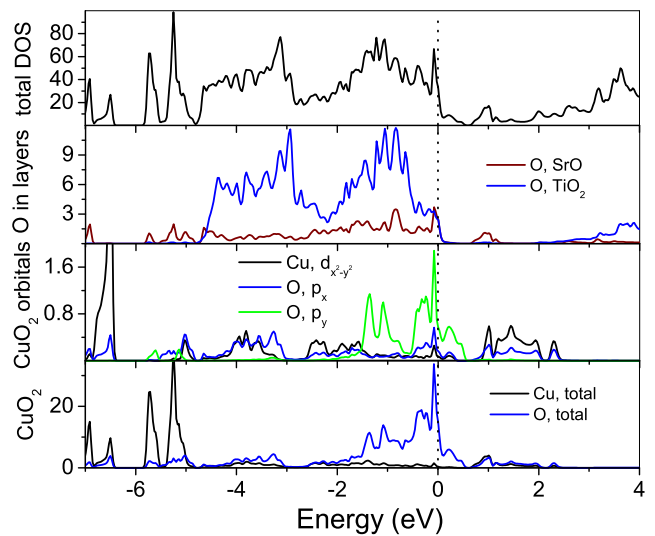}} \caption{Density of states of the
superlattice with CuO$_2$ deposited on SrTiO$_3$, terminated by SrO (LSDA+$U$
studies). The zero of energy is at Fermi level.} \label{fig5}
\end{figure}

In our theoretical studies, in order to focus on the effect of electronic
reconstruction, we have introduced a decoupling vacuum layer of $13$~\AA\
thickness between the CuO$_2$ surfaces in the superlattice of the same slab
geometry, as shown in Fig.~\ref{fig4}. Furthermore, we have optimized the
superlattice structure with a relaxed distance $\Delta=1.83$~\AA\ between the interface
CuO$_2$ and SrO which corresponds to a total energy minimum. The electronic
density of states calculated from LSDA+$U$ is shown in Fig.~\ref{fig5}. Here we see that the effect of
hole doping is more pronounced than in the case of the
YBa$_2$Cu$_2$O$_6$/SrTiO$_3$ sandwich: the energy gap between O $p$ and Cu $d$
states basically disappears and the Fermi level is located well below the top
of the valence band which characterizes a metallic state with hole carriers.
The hole charge is present in the CuO$_2$ planes, where it is hybridized
between the oxygen $p_x$ and $p_y$, and Cu $d$ orbitals. Also, there is clear
evidence for holes in the first SrO layer and in  the next TiO$_2$ layer of
SrTiO$_3$. In these layers, $E_F$ is also located below the top of the O $p$
bands.

\begin{figure}[t]
\epsfxsize=8.5cm {\epsffile{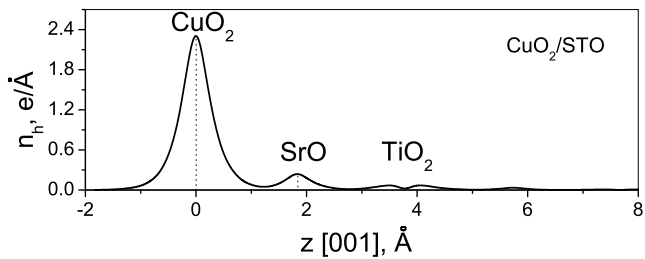}} \caption{Hole density distribution
in the interface planes of the CuO$_2$/SrTiO$_3$ superlattice. Here $z=0$
corresponds to the location of the lower CuO$_2$-plane.} \label{fig6}
\end{figure}

How is the hole charge redistributed near the interface? To provide more
details, we have calculated the density of holes in the planes nearest to the
interface. The results in Fig.~\ref{fig6} show that most of the charge
(about $80 \%$) is confined to the CuO$_2$ plane and a substantial amount of
hole density is located in the first SrO plane (about $11 \%$) and more distant
TiO$_2$ ($8 \%$) plane. Consequently, such heavy overdoping of CuO$_2$ should
completely exclude any possibility for superconducting state in the interface
unit cells of the cuprate films.

As the crystal structure of YBa$_2$Cu$_3$O$_{7-\delta}$ is closely compatible
with the perovskite SrTiO$_3$, all possible basic polar interface
configurations can be effectively reduced to the discussed two cases. In the
first case (A) with TiO$_2$-termination of SrTiO$_3$, the first interface layer
in the incomplete cell of YBa$_2$Cu$_3$O$_{7-\delta}$ is expected to be BaO in
order not to disturb the perovskite stacking. In the second case (B) with SrO
termination of SrTiO$_3$, the CuO$_2$ layer is the most compatible for a
continuous stacking (cf.~Fig.~\ref{fig4}). In both of these situations, the
interface polarity leads to a hole density in CuO$_2$ ranging from $\sim 0.05$
(case (A)) to $\sim 1.6$ (case (B)) holes per interface. In addition, due to
steps and different stacking modes at the interfaces \cite{bals}, one expects
rather a combination of cases (A) and (B). In fact, the latter implies the
formation of weak links with either connected or disconnected underdoped and
heavily-overdoped regions. For such possible interface configurations, the
direct consequence of the deduced pre-doping is a strong suppression of the
$T_c$ shift related to the hole injection--- or even a complete suppression of
the superconductivity in few unit cells of the cuprate films.

Up to now, the growth of high-quality YBa$_2$Cu$_3$O$_{7-\delta}$-films
terminated by complete unit cells remains to be a challenging task due to their
roughness caused by ionic compensation of the interface polarities. In order to
make a step towards perfect interfaces, where the hole injection would
completely determine the superconducting dome, we need to consider other
superconducting cuprates as possible candidates for non-polar interface. In
this context, a proposal for field effect experiments is to grow
Sr$_2$CuO$_2$Cl$_2$ (SCOC) or Ca$_2$CuO$_2$Cl$_2$ (CCOC) on SrTiO$_3$. These
systems can be perfectly cleaved between SrCl(CaCl)layers, and the deposition
on SrTiO$_3$ terminated by TiO$_2$ would result in a nonpolar interface stack
$\ldots$TiO$_2$/SrCl/CuO$_2$/SrCl/SrCl/$\ldots$ (for CCOC a similar stack with
Ca$\rightarrow$Sr). Our LSDA-calculations for such superlattices suggest an
insulating state. The lattice constants' mismatch ($a=b=3.96$~\AA\ in SCOC)
results in a slight increase ($V_{pd}/V_{pd}^{\rm bulk}=1.074$) of the $p$-$d$
hybridization integral $V_{pd}$ \cite{harrison}. Assuming the on-site Hubbard
coupling not to be affected by strains, the direct interface effect in
SCOC/SrTiO$_3$ is a renormalization of the parameters of the effective $t$--$J$
model \cite{zhang} ($t/t_{\rm bulk}=1.15$, $J/J_{\rm bulk}=1.33$) which would
only increase $T_c$, without changing the doping level. Such superlattices,
where the combination of chemical doping by Na and electrostatic hole injection
should not be affected by the interface pre-doping, would be ideal candidates
to probe the electrostatic field effect.

This work was supported through the DFG~SFB-484, BMBF~13N6918A, DAAD
D/03/36760, and by the Canadian funding agencies NSERC and CIAR. The research
was performed with infrastructure funded by CFI and British Columbia Knowledge
Development Fund (A Parallel Computer for Compact-Object Physics). Grants of computer time 
from the Leibniz-Rechenzentrum M\"unchen are gratefully acknowledged.


\begin{references}
\bibitem{multi} In the more general case when a multiplanar YBaCuO-layer terminated by CuO$_2$
is deposited on SrTiO$_3$, 1 extra hole is required to compensate the polarity. Despite the
high doping level, such a structure will be more stable than a single CuO$_2$ layer due to the
compensating microscopic fields.
\end{references}
\end{document}